\documentclass[10pt,twocolumn,pre,aps,showpacs,byrevtex,bibnotes]{revtex4}
\usepackage{graphicx}
\begin{document}

\title{Asymptotics of superstatistics}

\author{Hugo Touchette}

\email{htouchet@alum.mit.edu}

\affiliation{School of Mathematical Sciences, Queen Mary,
University of London, London E1 4NS, United Kingdom}

\author{Christian Beck}

\email{c.beck@qmul.ac.uk}

\affiliation{School of Mathematical Sciences, Queen Mary,
University of London, London E1 4NS, United Kingdom}

\date{\today}

\begin{abstract}

Superstatistics are superpositions of different statistics relevant for
driven nonequilibrium systems with spatiotemporal inhomogeneities of an
intensive variable (e.g., the inverse temperature). They contain Tsallis
statistics as a special case. We develop here a technique that allows us to
analyze the large energy asymptotics of the stationary distributions of
general superstatistics. A saddle-point approximation is developed which
relates this problem to a variational principle. Several examples are worked
out in detail.

\end{abstract}


\maketitle

\section{Introduction}

Many driven nonequilibrium systems exhibit complex dynamical behavior
characterized by spatiotemporal fluctuations of an intensive parameter
$\beta $ which may represent an inverse temperature, a chemical potential
(e.g., in a system with inhomogeneous concentrations), an effective friction
constant, the amplitude of a perturbing noise, or the local energy
dissipation, as in the case of turbulent flows. Although such systems do not
settle to equilibrium, their long-term behavior can often be described in
the spirit of equilibrium statistical mechanics by viewing them as
consisting of an ensemble of subsystems or ``cells'' to which are associated
different values of $\beta $. If there is local equilibrium in each cell, so
that statistical mechanics can be applied locally, and if the fluctuations
of $\beta $ evolve on a sufficiently large time scale, then in the long-term
run the entire system can be described by a mixture or superposition of
different Boltzmann factors having different values of $\beta $. Such a
mixture of various statistics has been termed a ``superstatistics''
\cite{beck-cohen} and has been the subject of various papers lately (see, e.g.,
Refs.~\cite{cohen, boltzmann-m, beck-su, touchette, sattin04, supergen, abestab,
abearbi, souza, souza2,luczka2000}). Many models based on the notion of
superstatistics have also been applied successfully to a variety of physical
problems, including Lagrangian \cite{beck03, reynolds03} and Eulerian
turbulence \cite{old-physica-d, beck-physica-d}, defect turbulence
\cite{daniels}, cosmic ray statistics \cite{beck-physica-a}, plasmas
\cite{sattin02}, statistics of wind velocity differences \cite{rapisarda}, and
econophysics \cite{ausloos, econo}.

What is common to all these problems is the experimental observation of
stationary distributions having ``fat'' tails. Such distributions fall
necessary outside the framework of ordinary statistical mechanics, but not
that of superstatistics. For example, if the random intensive parameter
$\beta $ in the various cells is taken to be distributed according to a
particular probability distribution, the $\chi ^2$-distribution, then the
corresponding superstatistics, obtained by integrating the Boltzmann factor
$e^{-\beta E}$ over all $\beta $, yields the nonextensive statistics of
Tsallis defined by the so-called $q$-exponential function
\cite{tsa1,tsa2,tsa3,abe}
\begin{equation}
e_q^{-\beta _0E}=[1+(q-1)\beta _0E]^{-1/(q-1)}.
\end{equation}
This particular statistics decays as a power-law for large energies $E$
rather than an exponential, as is the case for the ordinary Boltzmann
factor. In this sense, it is a fat-tailed statistics. The parameter
$\beta_0 $ above is related in the superstatistical model to the average inverse
temperature of the inhomogeneous system, whereas the so-called entropic
index $q$ relates to the variance of the $\beta $ fluctuations
\cite{wilk,prl}. It is worth mentioning that distributions having the form of a
$q$-exponential can be obtained formally by maximizing Tsallis' measure of
entropy subject to suitable constraints. Moreover, ordinary statistical
mechanics, which correspond in the superstatistics picture to the case where
there is no fluctuations in $\beta $, is recovered in the limit where
$q\to 1 $ \cite{tsa1,tsa2,tsa3,abe}.

For other distributions of the intensive parameter $\beta $, one ends up
with more general superstatistics. Generalized entropies, which are
analogs of the Tsallis entropies, can also be defined for these general
superstatistics \cite{abearbi,souza}, and generalized versions of
statistical mechanics can be constructed, at least in principle. It has been
shown that the corresponding generalized entropies are stable
\cite{abestab,souza2}.

In this paper we will briefly review the superstatistics concept, and then
analyze the asymptotic behavior of general superstatistics for large values
of the energy $E$. We will investigate how the properties of the function
$f(\beta )$, which represents the probability distribution of the intensive
variable $\beta $ in the various spatial cells, determine the asymptotic
decay rate of the generalized Boltzmann factor of the superstatistics. We
develop a saddle-point approximation technique which allows us to treat this
problem in full generality. Several examples will be worked out in detail to
show that the asymptotic decay rate of the stationary distributions
resulting from $f(\beta )$ can not only be a power-law, as in the case of
nonextensive statistics, but can also be an exponential of the square root
of the energy or, generally, a stretched exponential. We will discuss
universal aspects of the large energy asymptotics, thus complementing the
consideration in Ref.~\cite{beck-cohen} where universal aspects of the low-energy
behavior of general superstatistics were discussed. The large energy
asymptotics is of particular physical importance because the tails of the
observed distributions measured in various experiments (e.g., hydrodynamic
turbulence \cite{bodenschatz04}, plasmas \cite{carbone2000}, and granular
media \cite{vanzon2004}) can distinguish between the various possible types
of superstatistics.

\section{Superstatistics: basic concept}

Let us first briefly review the superstatistics concept as introduced in
Ref.~\cite{beck-cohen}. Consider a driven nonequilibrium system with
spatiotemporal fluctuations of an intensive parameter $\beta $, for example,
the inverse temperature. Locally, i.e.,\ in spatial regions (cells) where
$\beta $ is approximately constant, the system is described by ordinary
statistical mechanics, i.e., by\ an ordinary Boltzmann factor $e^{-\beta E}$,
where $E$ is an effective energy in each cell. In the long-term run, the
system is described by a spatiotemporal average over the fluctuating
$\beta $. In this way, one may define an effective Boltzmann factor $B(E)$ for the
whole system as
\begin{equation}
B(E)=\int_0^\infty f(\beta )e^{-\beta E}d\beta =\langle e^{-\beta E}\rangle ,
\end{equation}
where $f(\beta )$ is the normalized probability distribution describing the
$\beta $-fluctuations in the various cells. For so-called type-A
superstatistics, one normalizes this effective Boltzmann factor and obtains
the stationary, long-term probability distribution,
\begin{equation}
p(E)=\frac 1ZB(E),
\end{equation}
where 
\begin{equation}
Z=\int_0^\infty B(E)dE.
\end{equation}
For type-B superstatistics, the $\beta $-dependent normalization constant of
each local Boltzmann factor is included into the averaging process. In this
case, the invariant long-term distribution is given by
\begin{equation}
p(E)=\int_0^\infty f(\beta )\frac{e^{-\beta E}}{Z(\beta )}d\beta ,
\end{equation}
where $Z(\beta )$ is the normalization constant of the Boltzmann factor
$e^{-\beta E}$ for a given $\beta $. Both approaches can be easily mapped
into each other by considering a new probability density $\tilde{f}(\beta)
=c\cdot f(\beta )/Z(\beta )$, where $c$ is a normalization constant. One
immediately recognizes that type-B superstatistics with $f$ is equivalent to
type-A superstatistics with $\tilde{f}$.

As mentioned before, the fluctuations of the intensive parameter $\beta$
are spatiotemporal: they can be produced by either temporal changes
of the environment or by the movement of a test particle through
inhomogeneous spatial regions. The precise nature and behavior of $\beta$
in these situations can be quite varied. In our general description of
superstatistics, we have taken $\beta$ to be an inverse temperature
which varies randomly in time or in space, but $\beta$ can also represent,
say, a chemical potential that varies smoothly in space. In this way, one
may study superstatistical models where the fluctuations of $\beta$ are
caused by large-scale temperature gradients in a system
\cite{sattin04} or by a nonuniform chemical
potential describing inhomogeneous concentrations spread in space.

From a more dynamical perspective, a superstatistics can also be achieved
by considering
Langevin equations whose parameters fluctuate on a relatively large time
scale (the adiabatic regime \cite{note1}); see Ref.~\cite{prl} for details. For
turbulence applications, for example, one may consider a superstatistical
extension of the Sawford model of Lagrangian turbulence
\cite{beck03,reynolds03,sawford}. This model consists of suitable stochastic
differential equations for the position, velocity, and acceleration of a
Lagrangian test particle in the turbulent flow. The superstatistical
extension naturally enters due to the fact that the local energy dissipation
rate in turbulent flows is a random variable. Thus the parameters of the
Sawford model become random variables as well. These kinds of models well
reproduce experimentally measured turbulence data \cite{beck03,reynolds03}.

In the following we will use the notation of type-A superstatistics, keeping
in mind that we can always proceed to type-B superstatistics by replacing $f$
by $\tilde{f}$. We will restrict ourselves to positive values of $\beta $
and $E$ and will assume, in addition, that $f(\beta )$ is everywhere
differentiable and unimodal (single-bell-shaped curve).

\section{Low energy asymptotics}

\label{Slow}

We recall the low-energy asymptotics of superstatistics for reasons of
completeness; it was previously discussed in Ref.~\cite{beck-cohen}. Consider a
distribution $f(\beta )$ having mean $\left\langle \beta \right\rangle
=\beta _0$ and variance
\begin{equation}
\left\langle \beta ^2\right\rangle -\left\langle \beta \right\rangle
^2=\left\langle \beta ^2\right\rangle -\beta _0^2=\sigma ^2.
\end{equation}
Using the definition of $B(E)$, we can write
\begin{equation}
B(E)=\left\langle e^{-\beta E}\right\rangle =e^{-\beta _0E}\left\langle
e^{-(\beta -\beta _0)E}\right\rangle .
\end{equation}
Then, expanding in Taylor series the exponential term inside the expected
value, we obtain
\begin{equation}
B(E)=e^{-\beta _0E}\left( 1+\frac 12\sigma ^2E^2+\sum_{k=3}^\infty
\left\langle (\beta _0-\beta )^k\right\rangle \frac{E^k}{k!}\right) .
\label{taylor1}
\end{equation}
Thus, to second order in $E$, $B(E)$ must behave like
\begin{equation}
B(E)\sim e^{-\beta _0E}\left( 1+\frac 12\sigma ^2E^2\right)
\end{equation}
as $E\rightarrow 0$. This approximation represents the leading order
correction to ordinary statistical mechanics in our nonhomogeneous system
with temperature fluctuations for small values of the energy $E$. The
zeroth-order approximation to $B(E)$ corresponds, as is expected, to the
``pure'' Boltzmann statistics,
\begin{equation}
B(E)\sim e^{-\beta _0E},
\end{equation}
with inverse temperature $\beta _0$. It can be noted that these two
asymptotic results can also be considered to be valid approximations of
$B(E) $ in the limit where $\left\langle (\beta _0-\beta )^k\right\rangle
\rightarrow 0$ for all $k\geq 2$, that is essentially when $f(\beta
)\rightarrow \delta (\beta -\beta _0)$ (small fluctuations limit).

\section{High energy asymptotics}

\label{Shigh}

To find the high-energy asymptotics of $B(E)$, we use the fact that the
integral defining $B(E)$ has the form of a Laplace integral for
$E\rightarrow \infty $ \cite{bender1978}. In this limit, the integral can be
approximated by its largest integrand. This is the essence of Laplace
method, otherwise known as the saddle-point approximation
\cite{bender1978,murray1984}. The conditions of applicability of this
approximation method are basically the conditions that we have assumed
regarding the shape of $f(\beta )$ and its differentiability. Thus, putting
$B(E)$ in the form
\begin{equation}
B(E)=\int_0^\infty e^{-\beta E+\ln f(\beta )}d\beta ,
\end{equation}
we attempt to locate the largest integrand by finding the unique value of
$\beta $ which maximizes the exponent function,
\begin{equation}
\Phi (\beta ,E)=-\beta E+\ln f(\beta ),
\end{equation}
for any large enough energy value $E$. We call the value of $\beta $
maximizing $\Phi (\beta ,E)$ for fixed $E$ the \textit{dominant inverse
temperature} and denote it by $\beta _E$. The fact that $f(\beta )$ is
assumed to be unimodal ensures us that $\beta _E$ is unique, as required.
Indeed, observe that $\ln f(\beta )$ must be a concave function of $\beta $
if $f(\beta )$ is unimodal (see Fig.~1), and, in this case, the maximum of
$\Phi (\beta ,E)$ along the $\beta $-direction can only be attained at a
single point $\beta _E$ which is such that
\begin{equation}
E=[\ln f(\beta )]^{\prime }=\frac{f^{\prime }(\beta )}{f(\beta )}.
\label{leg1}
\end{equation}
Solving this equation for $\beta $, we find $\beta _E$ and thus write
\begin{eqnarray}
B(E) &\sim &e^{\Phi (\beta _E,E)}  \nonumber \\
&=&e^{-\beta _EE+\ln f(\beta _E)}  \nonumber \\
&=&f(\beta _E)e^{-\beta _EE}  \label{app1}
\end{eqnarray}
in the limit where $E\rightarrow \infty $. Note that this basic Laplace or
saddle-point approximation of $B(E)$ can be improved a bit more by
evaluating the integral defining $B(E)$ using a Gaussian approximation of
the integrand \cite{bender1978}. What results from this more refined
approximation is the following high-energy asymptotics:
\begin{equation}
B(E)\sim \frac{f(\beta _E)e^{-\beta _EE}}{\sqrt{-[\ln f(\beta _E)]^{\prime
\prime }}}  \label{app2}
\end{equation}
which differs from the Laplace approximation only by the square-root term
involving the curvature of $\ln f(\beta )$.

The Laplace approximation of $B(E)$ as well as its Gaussian corrected
version shown above are quite interesting from the physical point of view
because they show that the mixture of Boltzmann statistics defining $B(E)$
reduces at high energy $E$ to a ``pure'' Boltzmann statistics, just like in
the equilibrium situation, although now the Boltzmann statistics involves an
inverse temperature $\beta _E$ which depends on the energy $E$ considered.
This means that, for high values of $E$, the long-term, stationary behavior
of the nonequilibrium system considered is dominated by the equilibrium
behavior of a subset of cells having an inverse temperature equal or close
to $\beta _E$. How $\beta _E$ changes as a function of $E$ is determined
from the properties of $f(\beta )$. In the general case considered here,
where $\ln f(\beta )$ concave, we find that $\beta _E\rightarrow 0$ when
$E\rightarrow \infty $ (see Fig.~1). Thus in this case the large-energy
behavior of the generalized Boltzmann factors $B(E)$ is determined by the
small-$\beta $ (high temperature) behavior of the function $f(\beta )$.

To complete this section, note that the exponent function $\Phi (\beta _E,E)$
which enters in the asymptotics of $B(E)$ represents nothing but the
Legendre transform of $\ln f(\beta )$. The result of this transform is a
function of $E$ which can be thought of to represent, following the theory
of large deviations \cite{ellis1985}, an entropy function if we consider
that the function $\ln f(\beta )$ represents a free energy function. This
entropy function, however, is just a formal one and is unrelated to the
Tsallis entropies or other generalized entropies as defined in 
Refs.~\cite{tsa1,souza}.

\begin{figure}[t]
\centering
\includegraphics{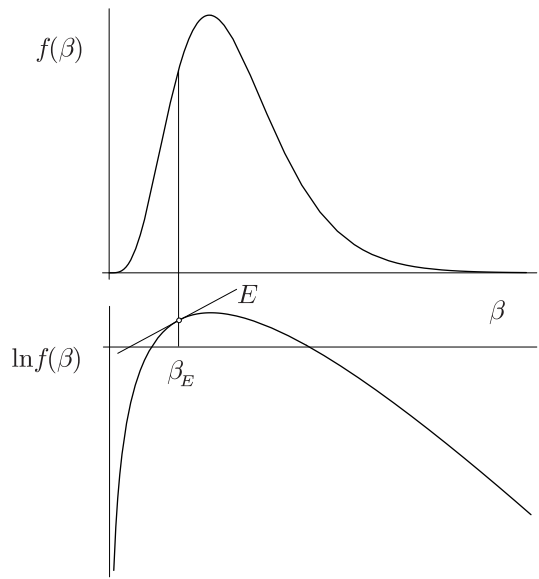}
\caption{Plot of a typical unimodal distribution $f(\beta)$ (top) and its logarithm (bottom).
The dominant inverse temperature $\beta_E$ is such that
$\ln f(\beta_E)$ has a slope equal to the energy value $E$.}
\label{uni1}
\end{figure}

\section{Examples}

We now consider a few cases of $\beta \rightarrow 0$ asymptotic behavior of
$f(\beta )$ and derive the corresponding asymptotic behavior of $B(E)$ for
$E\rightarrow \infty $ using the Laplace approximation of $B(E)$ corrected
with the square-root term, i.e., Eq.(\ref{app2}).

\subsection{Power-law tail}

Consider first an $f(\beta )$ with $f(\beta )\sim \beta ^\gamma $, $\gamma
>0 $ for $\beta \to 0$. An example of probability density having this
asymptotic form is the following $\chi ^2$ distribution for $\beta $
\cite{prl,wilk}:
\begin{equation}
f(\beta )=\frac 1{\Gamma (\frac n2)}\left( \frac n{2\beta _0}\right)
^{n/2}\beta ^{n/2-1}e^{-\frac{n\beta }{2\beta _0}},  \label{chi2}
\end{equation}
($\beta _0\geq 0$, $n>1$), which behaves as $f(\beta )\sim \beta ^{n/2-1}$
around $\beta =0$. Another example is the $F$-distribution
\cite{beck-cohen, sattin02},
\begin{equation}
f(\beta )=C\frac{\beta ^{\frac v2-1}}{(1+\frac{vb}w\beta )^{(v+w)/2}},
\end{equation}
where $v,w,b$ are parameters and $C$ is a normalization constant.

To find the asymptotic form of the superstatistics $B(E)$ corresponding to
the choice $f(\beta )\sim \beta ^\gamma $, we proceed to determine the value
of the dominant inverse temperature $\beta _E$ by solving Eq.(\ref{leg1}).
The solution here is $\beta _E=\gamma /E$, so that
\begin{eqnarray}
-\beta _EE+\ln f(\beta _E) &=&-\gamma +\gamma \ln \gamma -\gamma \ln E
\nonumber \\
&\sim &-\gamma \ln E
\end{eqnarray}
and
\begin{equation}
[\ln f(\beta _E)]^{\prime \prime }=-\frac \gamma {\beta _E^2}=-\frac{E^2}
\gamma \sim -E^2.
\end{equation}
Combining the two results in (\ref{app1}), we obtain
\begin{equation}
B(E)\sim \frac{e^{-\gamma \ln E}}E=E^{-\gamma -1}.
\end{equation}
Thus we see that power laws in $\beta $ for small $\beta $ imply a power
law in $E$ for large $E$, no matter what the rest of the distribution
$f(\beta )$ looks like. Comparing this asymptotic result with the
$q$-exponential distributions studied in nonextensive statistical mechanics
\cite{tsa1,tsa2,tsa3,abe}, which asymptotically decay as $E^{-1/(q-1)}$, we
see that 
\begin{equation}
\gamma +1=\frac 1{q-1}.  \label{here}
\end{equation}
Power-law superstatistics are physically relevant for many different
physical problems: e.g., defect turbulence \cite{daniels}, cosmic ray
statistics \cite{beck-physica-a}, and wind velocity measurements
\cite{rapisarda} among others.

\subsection{Exponential tail}

Consider now a density $f(\beta )$ having the asymptotic form $f(\beta )\sim
e^{-c/\beta }$, $c>0$, as $\beta \rightarrow 0$. An example is the inverse
$\chi ^2$ distribution
\begin{equation}
f(\beta )=\frac{\beta _0}{\Gamma (\frac n2)}\left( \frac{n\beta _0}2\right)
^{n/2}\beta ^{-n/2-2}e^{-\frac{n\beta _0}{2\beta }},
\end{equation}
($\beta _0>0$, $n>1$), which behaves, for $\beta \rightarrow 0$, as follows:
\begin{equation}
f(\beta )\sim \beta ^{-n/2-2}e^{-\frac{n\beta _0}{2\beta }}\sim e^{-\frac{
n\beta _0}{2\beta }}.
\end{equation}
This form of $f(\beta )$ was previously considered in the context of density
fluctuations in fusion plasma experiments \cite{sattin02} as well as
temperature fluctuations in perfect gases \cite{touchette}, and arises if
the temperature $T=(k_B\beta )^{-1}$ rather than $\beta $ itself is
$\chi ^2$-distributed.

For this example, the equation that we need to solve to find $\beta _E$ is
simply
\begin{equation}
[\ln f(\beta )]^{\prime }=\frac c{\beta ^2}=E,
\end{equation}
and so $\beta _E=\sqrt{c/E}$. As a result,
\begin{equation}
-\beta _EE+\ln f(\beta _E)=-2\sqrt{cE}
\end{equation}
and
\begin{equation}
B(E)\sim \frac{e^{-\beta _EE+\ln f(\beta _E)}}{\sqrt{-[\ln f(\beta
_E)]^{\prime \prime }}}\sim E^{-3/4}e^{-2\sqrt{cE}}
\end{equation}
using the fact that
\begin{equation}
[\ln f(\beta _E)]^{\prime \prime }=-\frac{2c}{\beta _E^3}=-\frac{2E^{3/2}}{
c^{1/2}}\sim -E^{3/2}.
\end{equation}
Here the effective Boltzmann factor $B(E)$ decays as an exponential of
$\sqrt{E}$. In particular, if $E=\frac 12u^2$ is a kinetic energy, one
obtains distributions that exhibit exponential tails in the velocity $u$
\cite{touchette}. This type of exponential behavior has been observed for
stationary distributions of the complex Ginzburg Landau equation \cite{chate},
as well as in fusion plasma experiments \cite{sattin02}.

\subsection{Stretched exponential tail}

As a generalization of the previous example, consider a density $f(\beta )$
which behaves as $f(\beta )\sim e^{-c\beta ^\delta }$, with $c>0$, $\delta
<0 $ as $\beta \to 0$. This particular form of stretched exponential implies
a high-energy behavior of $B(E)$ which also has the form of a stretched
exponential. Indeed, solving the differential equation satisfied by $\beta
_E $: 
\begin{equation}
[\ln f(\beta )]^{\prime }=-c\delta \beta ^{\delta -1}=c|\delta |\beta
^{\delta -1}=E,
\end{equation}
we find
\begin{equation}
\beta _E=\left( \frac E{c|\delta |}\right) ^{1/(\delta -1)}.
\end{equation}
With this value of $\beta _E$, the curvature of $\ln f(\beta )$ is
asymptotically evaluated as follows:
\begin{eqnarray}
[\ln f(\beta _E)]^{\prime \prime } &=&-c\delta (\delta -1)\beta _E^{\delta
-2}  \nonumber \\
&=&-c\delta (\delta -1)\left( \frac E{c|\delta |}\right) ^{(\delta
-2)/(\delta -1)}  \nonumber \\
&\sim &-E^{(\delta -2)/(\delta -1)}.
\end{eqnarray}
Similarly, the Legendre transform of $\ln f(\beta )$ is found to behave as
\begin{eqnarray}
-\beta _EE+\ln f(\beta _E) &=&\frac{E^{\delta /(\delta -1)}}{(c|\delta
|)^{1/(\delta -1)}}-\frac c{(c|\delta |)^{\delta /(\delta -1)}}E^{\delta
/(\delta -1)}  \nonumber \\
&=&\frac{E^{\delta /(\delta -1)}}{(c|\delta |)^{1/(\delta -1)}}\left(
1-\frac 1{|\delta |}\right)  \nonumber \\
&=&aE^{\delta /(\delta -1)}
\end{eqnarray}
as $E\rightarrow \infty $. From Eq.(\ref{app2}), we consequently obtain
\begin{equation}
B(E)\sim E^{(2-\delta )/(2\delta -2)}e^{aE^{\delta /(\delta -1)}}.
\end{equation}

Stretched exponentials are relevant for observed distributions in
hydrodynamic turbulence \cite{bodenschatz04}, plasma experiments
\cite{carbone2000}, as well as in granular media \cite{vanzon2004}. It is worth
pointing out that the idea of superposing exponential factors to obtain a
stretched exponential factor has been used previously by Palmer \textit{et
al.}~\cite{palmer1984} to model anomalous relaxation dynamics; see also
\cite{jou1990} for applications of the same idea in the context of dissipative
fluxes dynamics.

\subsection{Constant tail}

So far we have assumed that $f(\beta )\rightarrow 0$ as $\beta \rightarrow 0$.
We next consider a case where $f(\beta )$ goes to some constant as $\beta
\rightarrow 0$. For this case, the differential equation defining $\beta _E$
poses a problem because $[\ln f(\beta )]^{\prime }$ does not diverge as
$\beta $ approaches $0$. To define $\beta _E$, we must then resort to find
the largest value of $\Phi (\beta ,E)$ directly without the use of
derivatives. As an example, consider the case where the $\beta \to 0$
behavior of $f(\beta )$ is given by $f(\beta )\sim a$, with $a>0$. Then, 
\begin{eqnarray}
\beta _E &=&\arg \sup_\beta \{-\beta E+\ln a\}  \nonumber \\
&=&\arg \sup_\beta \{-\beta E\},
\end{eqnarray}
which implies that $\beta _E=0$ for all $E>0$. Consequently, $B(E)$ must
behave like a constant as $E\rightarrow \infty $, since 
\begin{equation}
-\beta _EE+\ln f(\beta _E)=\ln a.
\end{equation}
This is not a very interesting case physically because $B(E)$ is not
normalizable. Note that a constant asymptotics for $B(E)$ is also recovered
if $f(\beta )\sim ae^{-c\beta }$, $a>0$, $c>0$.

\subsection{Log-normal distribution}

For our final example, we consider the case where $\beta \in (0,\infty )$ is
distributed according to the log-normal density
\begin{equation}
f(\beta )=\frac a\beta \exp \left[ -c(\ln \beta -b)^2\right] ,
\end{equation}
where $a$, $b$, and $c$ are all positive constants. With this density, the
integral defining the stationary distribution $B(E)$ takes the form
\begin{equation}
B(E)=a\int_0^\infty \frac{d\beta }\beta e^{-\beta E}e^{-c(\ln \beta -b)^2}.
\end{equation}
This is equivalent to 
\begin{equation}
B(E)=a\int_{-\infty }^\infty e^{-c(y-b)^2-Ee^y}dy  \label{int2}
\end{equation}
using the change of variables $y=\ln \beta $. At this point, we proceed as
before to find the asymptotic behavior of $B(E)$ as $E\rightarrow \infty $
by locating the saddle-point $y_E$ of Eq.(\ref{int2}) which maximizes the
exponent function 
\begin{equation}
\Phi (y,E)=-c(y-b)^2-Ee^y
\end{equation}
over all real values of $y$. The exact solution of $y_E$ can be found to be
given by 
\begin{equation}
y_E=-W\left( \frac{Ee^b}{2c}\right) +b,
\end{equation}
where $W(x)$ is the Lambert or product-log function defined as the principal
branch solution of the equation $ye^y=x$. The Laplace approximation of $B(E)$
which results from this solution for $y_E$ is 
\begin{eqnarray}
\ln B(E) &\sim &\Phi (y_E,E)  \nonumber \\
&=&-cW\left( \frac{Ee^b}{2c}\right) ^2-Ee^{-W\left( \frac{Ee^b}{2c}\right)
+b}.
\end{eqnarray}

A more manageable asymptotics can be found analytically by expanding the
exponential term in $\Phi (y,E)$ around $y=0$ to first, second, or third
order in $y$, and then find the maximum of the corresponding approximation
of $\Phi (y_E,E)$ to obtain an approximation of the Laplace asymptotics of
$B(E)$ found above. This method is described in Ref.~\cite{holgate1989}, and
yields surprising good results already at third order in $y$ despite the
fact that $y_E\rightarrow -\infty $ when $E\rightarrow \infty $ (recall that
$\beta \rightarrow 0$ as $E\rightarrow \infty $ and that $y=\ln \beta $).

\section{Inverse problem}

The inverse transform of the Laplace transform defining $B(E)$ has the form

\begin{equation}
f(\beta )=\frac 1{2\pi i}\int_{c-i\infty }^{c+i\infty }B(E)e^{\beta E}dE,
\label{lap1}
\end{equation}
where $c$ is an arbitrary constant lying in the domain of definition of
$B(E) $. From this integral, we are in a position to predict the $\beta
\rightarrow 0$ or $\beta \rightarrow \infty $ behavior that $f(\beta )$
needs to have in order for $B(E)$ to behave according to some prescribed
form. This is the inverse problem of the previous sections, namely: given a
prescribed form for $B(E)$, what is the behavior of $f(\beta )$?

We will not reflect much on this inverse problem because the asymptotic
methods that can be used to solve it are exactly the same as those described
in the context of the direct problem. On the one hand, in the limit $\beta
\rightarrow 0$, we may proceed just as in Sec.~\ref{Slow} to expand the
exponential term $e^{\beta E}$ in (\ref{lap1}) to obtain a Taylor series for
$f(\beta )$:
\begin{eqnarray}
f(\beta ) &=&\frac 1{2\pi i}\int_{c-i\infty }^{c+i\infty }B(E)\left( 1+\beta
E+\frac{\beta ^2E^2}2+\cdots \right) dE  \nonumber \\
&=&a_0+a_1\beta +a_2\beta ^2+\cdots ,
\end{eqnarray}
where
\begin{equation}
a_k=\frac 1{2\pi i}\int_{c-i\infty }^{c+i\infty }B(E)\frac{(\beta E)^k}{k!}
dE.
\end{equation}
The coefficient $a_k$ can be computed by rotating the complex integral shown
above to the real line with the substitution $E\rightarrow iE$. On the other
hand, $\beta \rightarrow \infty $ asymptotics for $f(\beta )$ can be found
in just the same way as $E\rightarrow \infty $ asymptotics were found for
$B(E)$ using the Laplace method. At the level of the inverse Laplace
transform integral defining $f(\beta )$, the application of this
approximation method corrected with the Gaussian term yields
\begin{equation}
f(\beta )\sim \frac{e^{\beta E_\beta +\ln B(E_\beta )}}{\sqrt{-[\ln
B(E_\beta )]^{\prime \prime }}},
\end{equation}
for $\beta \rightarrow \infty $, where $E_\beta $ is now found by solving
the equation
\begin{equation}
[\ln B(E)]^{\prime }=-\beta
\end{equation}
for $E$. This represents a valid approximation of $f(\beta )$ at large
values of $\beta $ (low-temperature limit) provided, as was the case for
$f(\beta )$, that $B(E)$ is unimodal and differentiable. It is important to
notice that the energy value $E_\beta $ taken as function of $\beta $ is not
the inverse function of $\beta _E$. In the direct problem, $\beta _E$ is
found for $E\rightarrow \infty $, and in that limit we have seen that $\beta
_E\rightarrow 0$. For the asymptotics of the inverse problem, however, we
consider the limit $\beta \rightarrow \infty $.

\section{Conclusion}

We have analyzed the main asymptotic properties of general superstatistics
which are convex superpositions of Boltzmann exponential factors. The
saddle-point approximation method turns out to be useful to treat this
problem in full generality. In practice, our methods allow us to construct
simple superstatistical models that may underlie an experimentally measured
``fat tail'' distribution in a driven nonequilibrium system. In this context,
we have seen that what universally determines the form of the high energy
tail of observed distributions is the low $\beta $ (high temperature)
behavior of the mixing distribution $f(\beta )$ which defines at the bottom
the superstatistical model. The spectrum of possibilities where our results
can be applied is quite broad. It contains the power-law distributions of
nonextensive statistical mechanics as a special case, but it is also
relevant for stretched exponentials or lognormal superstatistics, as we have
demonstrated.

\begin{acknowledgments}

We thank Stefano Ruffo for useful comments on the
manuscript. H.T. was supported by the Natural Sciences and Engineering
Research Council of Canada and the Royal Society of London.

\end{acknowledgments}

\end{document}